\documentclass[10pt,final,twocolumn]{IEEEtran}
\usepackage{amssymb,amsfonts,amsthm}
\usepackage[cmex10]{amsmath}
\usepackage{cite}

\newtheorem{lem}{Lemma}
\newtheorem{thm}{Theorem}

\newtheorem{cor}{Corollary}
\newtheorem{defn}{Definition}
\newtheorem{remark}{Remark}
\newtheorem{assum}{Assumption}

\usepackage{graphicx,enumerate,subfigure}

\def\la{\lambda}

\def\vp{\varphi}
\def\ve{\varepsilon}
\def\vk{\varkappa}
\def\re{\mathop{Re}\nolimits}

\def\r{\mathbb R}

\def\be{\beta}

\def\be{\begin{equation}}
\def\ee{\end{equation}}
\def\ben{\begin{equation*}}
\def\een{\end{equation*}}

\newcommand{\dist}{{\rm dist}}
\newcommand{\sgn}{\mathop{\rm sign}\nolimits}
\newcommand{\dfb}{\stackrel{\Delta}{=}}
\newcommand{\tr}{\mathop{\rm Tr}\nolimits}

\begin{document}

\title{A guiding vector field algorithm for path following control of nonholonomic mobile robots}

\author{Yuri A. Kapitanyuk, Anton V. Proskurnikov and Ming Cao
\thanks{Yuri Kapitanyuk and Ming Cao are with the ENTEG institute at the University of Groningen, The Netherlands.
Anton V. Proskurnikov is with the Delft Center for Systems and Control at Delft University of Technology, Delft, The Netherlands, and also with
Institute for Problems of Mechanical Engineering of Russian Academy of Sciences (IPME RAS) and ITMO University, St. Petersburg, Russia. E-mail: \texttt{i.kapitaniuk@rug.nl, anton.p.1982@ieee.org, m.cao@rug.nl}}%
\thanks{Y. Kapitanyuk and M. Cao acknowledge the partial support from the European Research Council (ERCStG-
307207). The work of A. Proskurnikov is supported by NWO TTW, the Netherlands, under the project STW\#13712 ``From Individual Automated Vehicles to Cooperative Traffic Management - Predicting the benefits of automated driving through on-road human behaviour assessment and traffic flow models (IAVTRM)''.}%
}

\maketitle

\begin{abstract}                
In this paper we propose an algorithm for path-following control of the nonholonomic mobile robot based on the idea of the \emph{guiding vector field} (GVF). The desired path may be an arbitrary smooth curve in its implicit form, that is, a level set of a predefined smooth function. Using this function and the robot's kinematic model,
we design a GVF, whose integral curves converge to the trajectory. A nonlinear motion controller is then proposed which steers the robot along
such an integral curve, bringing it to the desired path. We establish global convergence conditions for our algorithm and demonstrate its applicability and performance by experiments with real wheeled robots.
\end{abstract}



\section{Introduction}

Many applications in industrial and mobile robots is built upon the functionality of following accurately a predefined geometric path \cite{handbook2005,handbook2008}. Path following is one of the central problems in automatic guidance of mobile robots, such as aerial vehicles~\cite{6712082}, underwater~\cite{Peymania2015} and surface~\cite{Fossen20142912} marine crafts, wheeled ground robots and autonomous cars~\cite{samson1992path,Ackerman1995}.
Although mathematical models of robots may differ, the principles of their path following control are similar and can be classified in several major categories.

One approach to steer to the desired path is to fix its time-parametrization; the path is thus treated as a function of time or, equivalently, the trajectory of some
reference point. Path steering is then reduced to reference-point following, which is a special case of the \emph{reference tracking} problem that has been extensively studied in control theory
and solved for a very broad class of nonlinear systems \cite{Isidori,Marconi}. An obvious advantage of the trajectory tracking approach is its applicability to a broad range of paths that can be e.g.
non-smooth or self-intersecting. However, this method has unavoidable fundamental limitations, especially when dealing with general nonlinear systems. As shown in \cite{Seron:1999}, unstable zero dynamics generally make it impossible to track a reference trajectory with arbitrary predefined accuracy; more precisely, the integral tracking error is uniformly positive independent of the controller design. Furthermore, in practice the robot's motion along the trajectory can be quite ``irregular'' due to its oscillations around the reference point. For instance, it is impossible to guarantee either path following at a precisely constant speed, or even the motion with a perfectly fixed direction.
Attempting to keep close to the reference point, the robot may ``overtake'' it due to unpredictable disturbances.

As has been clearly illustrated in the influential paper \cite{PFfnmpd}, these performance limitations of trajectory tracking can be removed by carefully designed \emph{path following} algorithms.
Unlike the tracking approach, path following control treats the path as a geometric curve rather than a function of time, dealing with its implicit equations or some time-free parametrization.
The algorithm thus becomes ``flexible to use a timing law as an additional control variable'' \cite{PFfnmpd}; this additional degree of freedom allows to maintain a constant forward speed or any other desired speed profile, which is extremely important e.g. for aerial vehicles, where the lifting force depends on the robot's speed.
The \emph{dynamic} controller, maintaining the longitudinal speed, is usually separated from ``geometric'' controller, steering the robot to the desired path.

A widely used approach to path-following, originally proposed for car-like wheeled robots \cite{samson1992path,micaelli1993trajectory},
assumes the existence of the \emph{projection}
point, that is, the closest point on a path, and the robot's capability to measure the distance to it (sometimes referred to as the ``cross-track error'').
The robot's mathematical model is represented in the Serret-Frenet frame, consisting of the tangent and normal vector to the trajectory at the projection point.
This representation allows to design efficient path following controllers for autonomous wheeled vehicles \cite{samson1992path,micaelli1993trajectory,Ackerman1995}
that eliminate the cross-track error and maintain the desired vehicle's speed along the path. Further development of this approach
leads to algorithms for the control of complex unmanned vehicles such as
cars with multiple trailers \cite{Samson1995} and agricultural tractors \cite{Matveev2013973}. Projection-based \emph{sliding mode} algorithms \cite{Ackerman1995,Matveev2013973}
are capable to cope with uncertainties, caused by the non-trivial geometry of the path, lateral drift of the vehicle and actuator saturations.

The necessity to measure the distance to the track imposes a number of limitations on the path-following algorithm.
Even if the nearest point is unique, the robot should either be equipped with special sensors \cite{Ackerman1995} or solve real-time optimization problems to find the cross-track error.
In general, the projection point cannot be uniquely determined when e.g. the robot passes a self-intersection point of the path
or its position is far from the desired trajectory. A possible way to avoid these difficulties has been suggested in \cite{1272868} and is referred to as the ``virtual target'' approach.
The Serret-Frenet frame, assigned to the projection point in the algorithm from \cite{samson1992path,micaelli1993trajectory}, can be considered as the
body frame of a \emph{virtual target vehicle} to be tracked by the real robot. A modification of this approach, offered in \cite{1272868}, allows the virtual target to have its own dynamics,
taken as one of the controller's design parameter. The design from \cite{1272868} is based on the path following controller from \cite{micaelli1993trajectory},
using the model representation in the Serret-Frenet frame and taking the geometry of the path into account. However, the controller from \cite{1272868}
implicitly involves target tracking since the frame has its own dynamics. Avoiding the projection problem, the virtual target approach thus inherits disadvantages of the usual target tracking.
In presence of uncertainties the robot may slow down and turn back in order to trace the target's position \cite{6712082}.

A guidance strategy of a human helmsman inspired another path following algorithm, referred to as the \emph{line-of-sight} (LOS) method \cite{6712082,1582226, Fossen20142912}
which is primarily used for air and marine crafts. Maintaining the desired speed of the robot, the LOS algorithm steers its heading along the LOS vector which
starts at the robot's center of gravity and ends at the \emph{target} point. This target is located ahead of the robot either on the path \cite{6712082} or on the line, tangent to the path
at the projection point \cite{Fossen20142912}. Unlike the virtual target approach, the target is always chosen at a fixed prescribed distance from the robot,
referred to as the \emph{lookahead distance}. The maneuvering characteristics substantially depend on the lookahead distance: the shorter distance is chosen, the more
``aggressive'' it steers. A thorough mathematical examination of the LOS method has been carried out in the recent paper \cite{Fossen20142912},
establishing the uniform semi-global exponential stability (USGES) property.

Differential-geometric methods for invariant sets stabilization \cite{fradkov1999nonlinear,Isidori,Marconi} have given rise to a broad class of ``set-based'' \cite{PBSS} path following algorithms.
Treating the path as a geometric set, the algorithm is designed to make it invariant and attractive (globally or locally).
Typically the path is considered as a set where some (nonlinear) output of the system vanishes, and thus the problem of its stabilization boils down to the output regulation problem.
To solve it, various linearization techniques have been proposed \cite{6355647, 7042843, 7002990,fradkov1999nonlinear, gn2013, lib:ifac_my, micnon,GillKulicNielsen2015}. For
stabilization of a closed \emph{strictly convex} curve an elegant passivity-based method has been established in \cite{PBSS}.

In this paper we develop a path following strategy, based on the idea of reference \emph{vector field}. Vector field algorithms are widely used in collision-free navigation and extremum seeking problems
\cite{PanagouKumar2014,HoyMatveevSavkin2015,matveev2017method,matveev2016extremum}; their efficiency in path following problems has been recently demonstrated in~\cite{lawrence2008lyapunov,4252175,6712082,6963394}. A vector field is designed in a way that its integral curves approach the path asymptotically. Steering the robot along the integral curves, the control algorithm drives it to the desired path. Unlike many path following algorithms, providing convergence only in a \emph{sufficiently small} vicinity of the desired path, the vector field algorithm guarantees convergence of any trajectory, which does not encounter the ``critical'' points where the vector field is degenerate. In particular, in any invariant domain without critical points the convergence to the path can be proved.

For holonomic robots described by a single or double integrator model a general vector-field algorithm for navigation along a general smooth curve in $n$-dimensional space has been discussed in \cite{5504176}. However, for more realistic nonholonomic vehicles the vector-field algorithms have been studied mainly for straight lines and circular paths \cite{lawrence2008lyapunov,4252175,6740857},
where they demonstrate better, in several aspects, performance compared to other approaches~\cite{6712082}.
Unlike \cite{lawrence2008lyapunov,4252175,6712082}, in this paper we propose and analyze rigorously a vector-field algorithm for guidance of a general nonholonomic robot along a general smooth planar path, given in its implicit form.

The paper is organized as follows. The path following problem is set up in Section~\ref{sec:setup}.
In Section~\ref{sec:field} we design the guiding vector field and discuss the properties of its integral curves.
Section~\ref{sec:algor} offers the path following control algorithm and establishes its main properties. This algorithm is practically validated by experiments with
wheeled robots, described in Section~\ref{sec:exper}. In Section~\ref{sec:discuss} we give a detailed comparison of our algorithm with several alternative approaches to path following controllers design and also discuss its possible extensions.

\section{Problem Setup}\label{sec:setup}

\begin{figure}
	\begin{center}
		\includegraphics[width=\columnwidth]{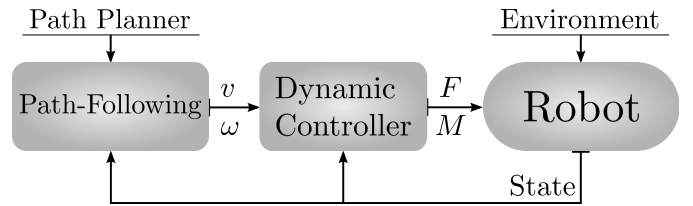}    
		\caption{The structure of motion control systems}
		\label{fig:0}
	\end{center}
\end{figure}

A widely used paradigm in path following control is to decompose the controller into an ``inner'' and an ``outer'' feedback loop~\cite{fossen1994guidance,beard2012small,tsourdos2010cooperative} as illustrated in Fig.~\ref{fig:0}. The ``inner'' \emph{dynamic} controller is responsible for maintaining the vector of generalized velocities (e.g. the longitudinal speed and turn rate)
by controlling the vector of forces and moments.
The design of the dynamic controller is based on the robot's mathematical model and may include rejection of external disturbances, e.g. adaptive drift compensators~\cite{Fossen2015Dubin}.
Assuming the dynamic controller to be sufficiently fast and precise, one may consider the speed and the turn rate to be new control inputs, describing thus the robot with a simpler \emph{kinematic} model. The path following algorithm is typically implemented in the ``outer'' \emph{kinematic} controller, steering the simplified kinematic model to the prescribed path.

The kinematic model of a mobile robot can be \emph{holonomic} or \emph{nonholonomic}. A holonomic robot is not restricted by its angular orientation and able to move in any direction,
as exemplified by helicopters, wheeled robots with omni-directional wheels~\cite{handbook2008} and fully actuated marine vessels at low speeds~\cite{1582226}. For nonholonomic robots only some
directions of motion are possible. The simplest model of this type is the \emph{unicycle}, which can move only in the longitudinal direction while the lateral motion is impossible. Examples of robots, reducing under certain conditions to the unicycle-type kinematics, include differential drive wheeled mobile robots~\cite{Samson1995,1359499}, fixed wing aircraft~\cite{lawrence2008lyapunov,4252175,beard2012small}, marine vessels at cruise speeds~\cite{fossen1994guidance} and car-like vehicles, whose rear wheels are not steerable~\cite{Samson1995,1359499,7042843}. The most interesting are unicycle models where the speed is restricted to be sufficiently high, making it impossible to reduce
the unicycle to a holonomic model via the feedback linearization~\cite{Oriolo2002}. The lift force of a fixed-wing UAV, the rudder's yaw moment of a marine craft and the turn rate of a car-like robot
depend on the longitudinal speed; at a low speed their manoeuvrability is very limited.

\subsection{The robot's model}

\begin{figure}[h]
	\begin{center}
		\includegraphics[width=\columnwidth]{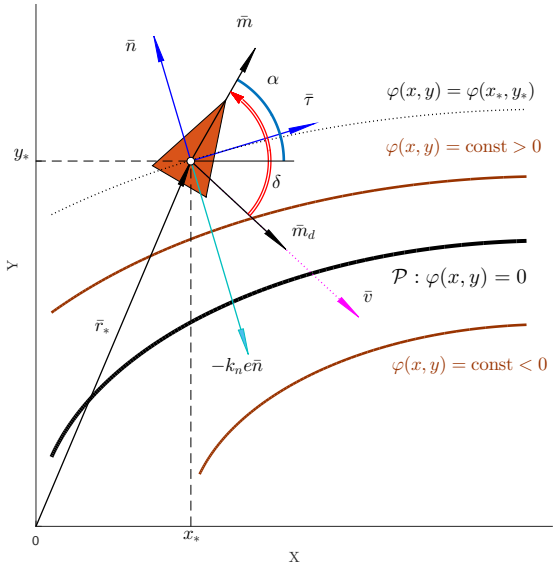}    
		\caption{The robot orientation and level sets of the function $\vp(x,y)$}
		\label{fig:1}
	\end{center}
\end{figure}

In this paper, we consider the unicycle-type model where the longitudinal velocity $u_r>0$ is a predefined constant
\begin{equation}\label{eq:robot}
\begin{gathered}
\dot {\bar r}=\begin{bmatrix}
\dot x\\
\dot y
\end{bmatrix}=u_r\bar m(\alpha)\in\r^2,\quad \bar m(\alpha)=\begin{bmatrix}
\cos\alpha\\
\sin\alpha
\end{bmatrix},\\
\dot{\alpha}=\omega.
\end{gathered}
\end{equation}
Here $\bar{r}$ is the position of the robot's center of gravity $C$
in the inertial Cartesian frame of reference $0XY$, $\alpha$ is the robot's orientation in this frame and $\bar m(\alpha)$ is the unit orientation vector (see Fig.~\ref{fig:1}).
The only control input to the system~\eqref{eq:robot} is the angular velocity $\omega$, and hence the system is underactuated.

\subsection{The description of the desired path}

 The desired curvilinear path $\mathcal{P}$ is the zero set of a function $\varphi\in C^2(\r^2\to\r)$, i.e. it is described by the implicit equation
\begin{equation}\label{eq:phi}
\mathcal{P}\dfb\{(x,y): \varphi(x,y)=0\}\subset\r^2.
\end{equation}
The same curve $\mathcal P$ may be represented in the implicit form \eqref{eq:phi} in many ways. The principal restrictions imposed by our approach is \emph{regularity}~\cite{fradkov1999nonlinear}: in some vicinity of $\mathcal P$ one has
\begin{equation}\label{eq:non-degen}
\bar n(x,y)\dfb\nabla\varphi(x,y)=\left[
\begin{array}{cc}
\frac{\partial \varphi(x,y)}{\partial x} ; \frac{\partial \varphi(x,y)}{\partial y}
\end{array}
\right]^{\top}\neq 0.
\end{equation}

As illustrated in Fig.~\ref{fig:1}, the plane $\r^2$ is covered by the disjoint \emph{level sets} of the function $\vp$, namely
the sets where $\vp(x,y)=c=const$. If~\eqref{eq:non-degen} holds, then the vector $\bar n(x,y)$ is the \emph{normal} vector to the corresponding level set at the point $(x,y)$.
The path $\mathcal P$ is one of the level sets, corresponding to $c=0$; the value $\varphi(x(t),y(t))$ can be considered as a (signed) ``distance'' from the robot to the path (differing, as usual, from the Euclidean distance), or the
\emph{tracking error}~\cite{fradkov1999nonlinear,gn2013}. More generally, choosing an arbitrary strictly increasing function $\psi\in C^1(\r\to\r)$ with $\psi(0)=0$ (and thus $\psi(s)s>0$ for any $s\ne 0$), one may define the tracking error as follows
\begin{equation}
\label{eq:e}
e(x,y)\dfb\psi[\varphi(x,y)]\in\r.
\end{equation}
By definition, $e=0$ if and only if $(x,y)\in\mathcal P$.
Our goal is to design a path following algorithm, i.e. a causal feedback law $(x(\cdot),y(\cdot),\alpha(\cdot))\mapsto \omega(\cdot)$, which eliminates the tracking error
$|e(t)|\xrightarrow[t\to\infty]{} 0$, bringing thus the robot to the predefined path $\mathcal P$.
Upon reaching the desired path, the algorithm should ``stabilize'' the robot on it, which means that
the robot's heading is steered along the tangent vector to the path.

The mapping $\psi(\cdot)$ in \eqref{eq:e} is a free parameter of the algorithm. Formally one can get rid of this parameter, replacing $\vp$ by the composition $\psi\circ\vp$. However,
it is convenient to distinguish between the path-defining function $\vp(x,y)$ and the tracking error, depending on the choice of $\psi(s)$.
The path representation is usually chosen as simple as possible: for instance, dealing with a straight line, it is natural to choose linear $\vp(x,y)$, while the circular path is naturally
described by $\vp(x,y)=(x-x_0)^2+(y-y_0)^2$. At the same time, some mathematical properties of the algorithm, in particular, the region where convergence of the algorithm is guaranteed,
depend on the way the tracking error is calculated. As will be discussed, it may be convenient to choose $\psi(\cdot)$ bounded with $|\psi'(s)|\xrightarrow[|s|\to\infty]{}0$,
for instance, $\psi(s)=\arctan s^p$ or $\psi(s)=|s|^p\sgn s/(1+|s|^p)$ with $p\ge 1$. The mentioned two functions, as well as the simplest function $\psi(s)=s$ satisfy the condition,
which is henceforth assumed to be valid
\be\label{eq:sup}
\sup_{u\in\r}\psi'(\psi^{-1}(u))<\infty.
\ee

\subsection{Technical assumptions}

Henceforth the following three technical assumptions are adopted, excluding some ``pathological'' situations.
Our first assumption enables one to use of the tracking error $e(x,y)$ as a ``signed distance''
to the path $\mathcal P$ and implies that the \emph{asymptotic} vanishing of the error $e(\bar r(t))\xrightarrow[t\to\infty]{} 0$ entails the convergence to the path in the usual Euclidean metric\footnote{As usual, the distance from a point $\bar r_0$ to the path $\mathcal P$
$A$ is $\dist(\bar r_0,\mathcal P)=\inf\{|\bar r_0-\bar r|:\,\bar r\in \mathcal P\}$. More generally, the distance between sets $A,B$ is defined as
$\dist(A,B)=\inf\{|\bar r_1-\bar r_2|:\,\bar r_1\in A,\bar r_2\in B$\}.} $\dist(\bar r(t),\mathcal P)\to 0$.
\begin{assum}\label{ass:non-degen}
For an arbitrary constant $\vk>0$ one has
\be\label{eq:separated1}
\begin{split}
\inf\{|e(\bar r)|:\dist(\bar r,\mathcal P)\ge\vk\}>0.
\end{split}
\ee
\end{assum}

Our second assumption provides the regularity condition~\eqref{eq:non-degen} in a sufficiently small vicinity of $\mathcal P$.
\begin{assum}\label{ass:positive}
The set of \textbf{critical points}, where $\bar n$ vanishes,
\ben
\mathcal C_0\dfb\{(x,y)\in\r^2:\nabla\vp(x,y)=0\},
\een
is separated from $\mathcal P$ by a positive distance $\dist(\mathcal C_0,\mathcal P)>0$.
\end{assum}
In the most typical cases where either $\mathcal P$ is a closed curve or $\mathcal C_0$ is compact, Assumption~\ref{ass:positive} boils down to the condition $\mathcal C_0\cap \mathcal P=\emptyset$.
Our final assumption is similar in spirit to Assumption~\ref{ass:non-degen} and guarantees that
the asymptotic vanishing of the normal vector $\bar n(r(t))\xrightarrow[t\to\infty]{} 0$ is possible only along a trajectory, converging to $\mathcal C_0$, i.e. $\dist(\bar r(t),\mathcal C_0)\to 0$.
\begin{assum}\label{ass:regularity}
For an arbitrary constant $\vk>0$ one has
\be\label{eq:separated1}
\begin{split}
\inf\{|\bar n(\bar r)|:\dist(\bar r,\mathcal C_0)\ge\vk\}>0.
\end{split}
\ee
\end{assum}

Assumption~\ref{ass:regularity} implies the following useful technical lemma.
\begin{lem}\label{lem:n}
Consider a Lipschitz vector-function $\bar r:[0;\infty)\to\r^2$ such that $\bar r(t)$ \emph{does not} converge to $\mathcal C_0$ as $t\to\infty$, that is, $\limsup\limits_{t\to\infty}d(t)>0$ where $d(t)=\dist(\bar r(t),\mathcal C_0)$.
Then $\int_0^{\infty}|\bar n(\bar r(t))|^pdt=\infty$ for any $p>0$.
\end{lem}
\begin{IEEEproof}
It can be easily noticed that $d(\cdot)$ is a Lipschitz function, since as can be easily shown, $|d(t_1)-d(t_2)|\le |\bar r(t_1)-\bar r(t_2)|\le M|t_1-t_2|\,\forall t_1,t_2\ge 0$, where $M>0$ is the Lipschitz constant for $\bar r(\cdot)$. Since $\limsup\limits_{t\to\infty}d(t)>0$,
a number $\ve>0$ and a sequence $t_k\to\infty$ exist such that $d(t_k)\ge 2\ve$ and, therefore, $d(t)\ge\ve$ for $t\in \Delta_k=[t_k;t_k+M^{-1}\ve]$. Passing to subsequences, one may assume without loss of generality that
$t_k+M^{-1}\ve<t_{k+1}$ and thus the intervals $\Delta_k$ are disjoint. Assumption~\ref{ass:regularity} implies that for sufficiently small $c>0$ one has $|\bar n(\bar r(t))|\ge c$ whenever $t\in\Delta_k$. Hence $\int_{\Delta_k}|\bar n(\bar r(t))|^pdt\ge \ve M^{-1}c^p$ and therefore $\int_0^{\infty}|\bar n(\bar r(t))|^p\,dt=\infty$.
\end{IEEEproof}

\section{The Guiding Vector Field and Its Properties}\label{sec:field}

In this section, we construct the Guiding Vector Field (GVF) to be used in the path following control algorithm.
We show that the integral curves of this field lead either to the desired path $\mathcal P$ or to the critical set $\mathcal C_0$. Furthermore, we give efficient criteria, ensuring that the integral curves of the second type are either absent or cover a set of zero measure on $\mathbb{R}^2$.

Besides the normal vector $\bar n(x,y)=\nabla\vp(x,y)$, at each point we consider the \emph{tangent} vector to
the level set
\be\label{eq:tau}
\bar\tau(x,y) =E\bar n(x,y),\quad E=\left[
\begin{array}{cc}
0 & 1\\ -1 & 0
\end{array}
\right].
\ee
If the point is regular~\eqref{eq:non-degen}, the basis $(\bar\tau,\bar n)$ is right-handed
oriented\footnote{In the subsequent constructions, one may replace $\bar{\tau}$ with $-\bar{\tau}$ and $E$ to $-E$, which causes the change of the path following direction.} (Fig.~\ref{fig:1}).
Our goal is to find such a vector field $\bar v(x,y)$, where the absolute tracking error $|e|$ is decreasing along each of its integral curves (unless $e=0$), and the curves starting on $\mathcal P$ do not leave it. We define this vector field by
\begin{equation}\label{eq:v}
\bar v(x,y)\dfb\bar\tau(x,y) - k_n e(x,y)\bar{n}(x,y).
\end{equation}

The integral curves of the vector field~\eqref{eq:v} correspond to the trajectories of the autonomous differential equation
\be\label{eq:sys-v}
\frac{d}{dt}\bar \xi(t)=\bar v(\bar \xi(t))\in\r^2,\quad t\ge 0.
\ee

\subsection{Properties of the integral curves}

We notice first that the vector field $\bar v$ is $C^1$-smooth, which implies the \emph{local} existence and uniqueness of solutions of the system~\eqref{eq:sys-v}. Obviously, any point $(x_0,y_0)\in\mathcal C_0$ corresponds to
the \emph{equilibrium} of~\eqref{eq:sys-v}, being a trivial single-point integral curve.
The uniqueness property implies that non-constant integral curves are free of the critical points; the corresponding solutions, however, may converge to $\mathcal C_0$ \emph{asymptotically}. In general, a solution may also escape to infinity in finite time.

The following lemma establishes the principal \emph{dichotomy} property of the integral curves, stating that any curve leads either to the desired path $\mathcal P$ or to the critical set $\mathcal C_0$.
\begin{lem}\label{lem:converge}
Let $\bar \xi(t)$, $t\in [0;t_*)$ be a maximally prolonged solution to~\eqref{eq:sys-v}. Then two situations are possible
\begin{enumerate}
\item either $\dist(\bar\xi(t),\mathcal P)\xrightarrow[t\to t_*]{}0$, that is, the solution converges to the desired path,
\item or $t_*=\infty$ and $\dist(\bar\xi(t),\mathcal C_0)\xrightarrow[t\to \infty]{}0$.
\end{enumerate}
\end{lem}
\begin{IEEEproof}
The proof is based on the Lyapunov function
\begin{equation}\label{eq:V-lyap}
V(x,y) = \frac{1}{2}e(x,y)^2\ge 0.
\end{equation}
A straightforward computation shows that $\nabla e(\bar\xi)=\psi'(\vp(\bar \xi))\,\nabla\vp(\bar \xi)=\psi'(\psi^{-1}(e(\bar\xi)))|\bar n(\bar\xi)|$
and hence $V$ is non-increasing along the trajectory $\bar\xi(t)$ since its derivative is
\begin{equation}\label{eq:dvt-1}
\begin{aligned}
\dot{V}(\bar \xi) &=e(\bar \xi)\psi'(\psi^{-1}(e(\bar \xi)))\bar n^{\top}(\bar \xi)\bar v(\bar \xi)\stackrel{\eqref{eq:v}}{=}\\&\stackrel{\eqref{eq:v}}{=}-k_n e(\bar \xi)^2\psi'(\psi^{-1}(e(\bar \xi)))|\bar n(\bar \xi)|^2\le 0.
\end{aligned}
\end{equation}
In particular, there exists the limit $e_*=\lim\limits_{t\to t_*}|e(\bar \xi(t))|\ge 0$. If $e_*=0$ then, due to Assumption~\ref{ass:non-degen}, statement 1) holds: the solution converges to $\mathcal P$. Suppose now that $e_*>0$. We are going to prove that statement 2) is valid. Notice first that $\psi'(\psi^{-1}(e(\bar \xi(t)))$ is uniformly bounded and positive; the same holds for $V(\bar\xi(t))$. The equality~\eqref{eq:dvt-1} thus implies that $\int_0^{t_*}|\bar n(\bar\xi(t))|^2dt<\infty$. Since $|v|^2=(1+k_n^2e^2)|\bar n|^2$,~\eqref{eq:sys-v} implies that
$\int_0^{t_*}|\dot{\bar\xi}(t))|^2dt<\infty$. If one had $t_*<\infty$, the Cauchy-Schwartz inequality would imply that
\[
|\bar \xi(T)-\bar \xi(0)|^2=\left(\int_0^{T}|\dot{\bar \xi}(t)|dt\right)^2\le t_*\int_0^{t_*}|\dot{\bar \xi}(t)|^2dt\;\;\forall T<t_*,
\]
which contradicts the assumption that $\bar\xi(t)$ is a maximally prolonged solution, escaping to infinity as $t\to t_*$. Therefore, $t_*=\infty$; statement~2 now follows from Lemma~\ref{lem:n}.
\end{IEEEproof}

A natural question arises on how ``large'' the set of trajectories is, converging to the critical set $\mathcal C_0$. In many practical examples, this set is \emph{finite}.
This holds, in particular if $\vp(x,y)\ne 0$ when $|x|+|y|$ is sufficiently large and its \emph{Hessian matrix}
\be\label{eq:hessian}
H(x,y)=H(x,y)^{\top}=
\begin{bmatrix}
\frac{\partial^2}{\partial x^2}\varphi(x,y) & \frac{\partial^2}{\partial x\partial y}\varphi(x,y)\\
\frac{\partial^2}{\partial x\partial y}\varphi(x,y) & \frac{\partial^2}{\partial y^2}\varphi(x,y)
\end{bmatrix}
\ee
is sign-definite at any point $(x_0,y_0)\in \mathcal C_0$, i.e. either $H(x_0,y_0)>0$ or $H(x_0,y_0)<0$. In this situation $\mathcal C_0$ is bounded (and thus compact) and all its points are isolated, which implies that $\mathcal C_0$ is finite. If the set $\mathcal C_0$ is finite, the convergence $\dist(\bar\xi(t),\mathcal C_0)\xrightarrow[t\to \infty]{}0$, obviously, means that the solution converges to a critical point: $\bar\xi_*\dfb\lim\limits_{t\to\infty}\bar\xi(t)\in\mathcal C_0$.
\begin{defn}
Let $\bar\xi_*\in\mathcal C_0$ be an equilibrium point of~\eqref{eq:sys-v}. The \emph{stable manifold} of $\bar\xi_*$, denoted by $W(\bar\xi_*)$, is the set of all points $\bar\xi_0$ such that
the solution of~\eqref{eq:sys-v}, starting at $\xi(0)=\xi_0$, exists for all $t\ge 0$ and $\bar\xi_*\dfb\lim\limits_{t\to\infty}\bar\xi(t)$.
\end{defn}

We will use the following corollary of the central manifold theorem (see e.g. Theorem~4.1 and Proposition 4.1 in~\cite{Monzon:2009}).
\begin{lem}\label{lem:monzon}
If both eigenvalues of the Jacobian matrix $J(\bar\xi_*)=\frac{\partial}{\partial\bar\xi}\bar v(\bar\xi_*)$ are strict unstable $\re\la_{1,2}J(\bar\xi_*)>0$, then $W(\bar\xi_*)=\{\bar\xi_*\}$.
If $J(\bar\xi_*)$ has at least one strictly unstable eigenvalue, then $W(\bar\xi)$ is a set of zero measure.
\end{lem}
Lemma~\ref{lem:monzon} in turn has the following important corollary.
\begin{cor}\label{cor:almost-all}
Let the set $\mathcal C_0$ be finite and for any of its points $\bar\xi_*\in\mathcal C_0$ the matrix $e(\bar\xi_*)H(\bar\xi_*)$ has a \emph{negative}\footnote{Recall that $H$ is a symmetric matrix, so its eigenvalues are real} eigenvalue. Then the maximally prolonged solution of~\eqref{eq:sys-v} (possibly, existing on finite interval only) converges to $\mathcal P$ for almost all initial conditions $\bar\xi(0)$. If $e(\bar\xi_*)H(\bar\xi_*)<0$ for any $\xi_*\in\mathcal C_0$, this convergence takes place whenever $\bar\xi(0)\not\in\mathcal C_0$.
\end{cor}
\begin{IEEEproof}
A straightforward computation shows that if $\nabla\vp(\bar\xi_*)=0$ then $J(\bar\xi_*)=(E-k_ne(\bar\xi_*))H(\bar\xi_*)$ and
\[
\begin{gathered}
\tr J(\bar\xi_*)=-k_ne(\bar\xi_*)\tr H(\bar\xi_*),\\
\det J(\bar\xi_*)=(1+k_n^2e(\bar\xi_*)^2)\det H(\bar\xi_*).
\end{gathered}
\]
It can be easily shown that a symmetric $2\times 2$ matrix $M$ is non-negatively definite $M\ge 0$ if and only $\det M=\la_1(M)\la_2(M)\ge 0$ and $\tr M=\la_1(M)+\la_2(M)\ge 0$. Here $\la_i(M)$ stand for the eigenvalues of $M$. Since $eH(e)$ is \emph{not} non-negatively definite, either $\det J<0$ or $\tr J>0,\det J\ge 0$. In both situations, the matrix $J(\bar\xi_*)$ has at least one strictly unstable eigenvalue. Furthermore, if $eH(e)<0$ then $\tr J>0,\det J>0$, i.e. both eigenvalues of $J$ are strictly unstable. The statement now follows from Lemma~\ref{lem:monzon}.
\end{IEEEproof}

\subsection{The GVF and the ``ideal'' motion of the robot}

The main idea of the path following controller, designed in the next section, is to steer the robot to the integral curve of the field~\eqref{eq:v}. In other words, when the robot is passing a point $\bar r=(x,y)\in\r^2$, its desired orientation is
\be\label{eq:md}
\bar m_d(x,y)=\frac{1}{|\bar v(x,y)|}\bar v(x,y).
\ee
The field of unit vectors $\bar m_d(x,y)$, henceforth referred to as the \emph{guiding vector field} (GVF),
is defined at any regular point $(x,y)$, where $\bar n\ne 0$ and thus $|\bar v|=\sqrt{1+k_n^2e^2}|\bar n|\ne 0$. Fig.~\ref{fig:1} illustrates the relation between the vectors $\bar r$, $\bar m$, $\bar\tau$, $\bar n$, $\bar v$, $\bar m_d$.

Consider the desired motion of the robot, ``ideally'' oriented along the integral curves of the GVF at any point. Its position vector $\bar r(t)$ obeys the differential equation
\be\label{eq:sys-md}
\dot{\bar r}(t)=u_r\bar m_d(\bar r(t)),\quad t\ge 0.
\ee
The following lemma is dual to Lemma~\ref{lem:converge}.
\begin{lem}\label{lem:converge1}
Let $\bar r(t)$, $t\in [0;t_*)$ be a maximally prolonged solution to~\eqref{eq:sys-md}. Then two situations are possible
\begin{enumerate}
\item either $\dist(\bar r(t),\mathcal C_0)\xrightarrow[t\to t_*]{}0$,
\item or $t_*=\infty$ and $\dist(\bar r(t),\mathcal P)\xrightarrow[t\to \infty]{}0$ that is, the solution converges to the desired path;
\end{enumerate}
\end{lem}
\begin{IEEEproof}
If $t_*<\infty$ then the limit $\bar r_*=\bar r(t_*-0)=\bar r(0)+\int_0^{t_*}\bar m_d(\bar r(t))dt$ exists and, since the solution is maximally prolonged, one obviously has $|\bar n(\bar r_*)|=0$, i.e. $\bar r_*\in\mathcal C_0$ and statement~1 holds. The case of $t_*=\infty$ is considered similar to the proof of Lemma~\ref{lem:converge}. Introducing the Lyapunov function~\eqref{eq:V-lyap}, its derivative is shown to be
\begin{equation}\label{eq:dvt-1+}
\begin{aligned}
\dot{V}(\bar r) &=u_re(\bar r)\psi'(\psi^{-1}(e(\bar r)))\bar n^{\top}(\bar r)\bar m_d(\bar r)\stackrel{\eqref{eq:v}}{=}\\&\stackrel{\eqref{eq:v}}{=}-\underbrace{\frac{u_rk_ne(\bar r)^2}{\sqrt{1+k_n^2e(\bar r)^2}}\psi'(\psi^{-1}(e(\bar r)))}_{\Psi(e(\bar r))}|\bar n(\bar r)|\le 0.
\end{aligned}
\end{equation}
There exists the limit $e_*=\lim_{t\to\infty}|e(\bar r(t))|$. If $e_*=0$, statement~2 holds thanks to Assumption~\ref{ass:non-degen}. Otherwise, $e_*>0$ and $\Psi(e(\bar r(t)))$ is uniformly positive.
In view of~\eqref{eq:dvt-1+}, one has $\int_0^{\infty}|\bar n(\bar r(t))|dt<\infty$. Lemma~\ref{lem:n} entails now statement 1.
\end{IEEEproof}

\begin{remark}
The equation~\eqref{eq:dvt-1+} implies that $\dot V\approx -k_nV|\bar n|\theta(e)$, where $\theta(e)\to \theta_0>0$ as $e\to 0$. Assumptions~\ref{ass:positive} and~\ref{ass:regularity} imply that $|\bar n(r)|\ge\vk$ as $e\approx 0$. Therefore, the desired path $\mathcal P$ is \emph{locally} exponentially attractive: if $|e(0)|\le\ve$, where $\ve$ is sufficiently small, then $|e(t)|\le e^{-k_nu_r\beta t}|e(0)|$, where $\beta>0$ is a constant, depending on $\psi(\cdot)$ and the normal vector $\bar n(x,y)$ in the vicinity of $\mathcal P$.
The coefficient $k_n$ is responsible for the attraction to $\mathcal P$. In the limit case $k_n=0$ the path $\mathcal P$ is not asymptotically stable; the higher $k_n>0$ one chooses, the stronger is the attraction to the path $\mathcal P$ in its small vicinity.
\end{remark}

Obviously, the integral curves of~\eqref{eq:sys-md} are in one-to-one correspondence with non-equilibrium integral curves of~\eqref{eq:sys-v}. Corollary~\ref{cor:almost-all} can now be reformulated as follows.
\begin{cor}\label{cor:almost-all+}
Let the set $\mathcal C_0$ be finite and for any of its points $\bar\xi_*\in\mathcal C_0$ the matrix $e(\bar\xi_*)H(\bar\xi_*)$ has a negative eigenvalue. Then for almost all initial conditions $\bar r(0)$ the solutions of~\eqref{eq:sys-md} can be prolonged up to $\infty$ and converge to $\mathcal P$. If $e(\bar\xi_*)H(\bar\xi_*)<0$ for any $\xi_*\in\mathcal C_0$, this holds for any $\bar r(0)\not\in\mathcal C_0$.
\end{cor}

\begin{figure}
	\begin{center}
		\includegraphics[width=0.75\columnwidth]{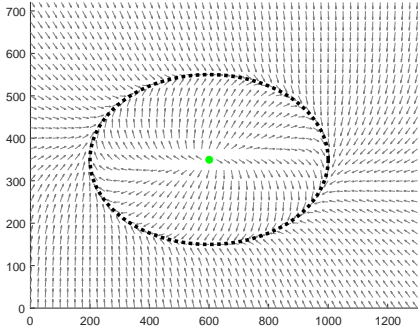}    
		\caption{The GVF for an elliptic path}
		\label{fig:ell-xy-field}
	\end{center}
\end{figure}

\begin{figure}
	\begin{center}
		\includegraphics[width=0.75\columnwidth]{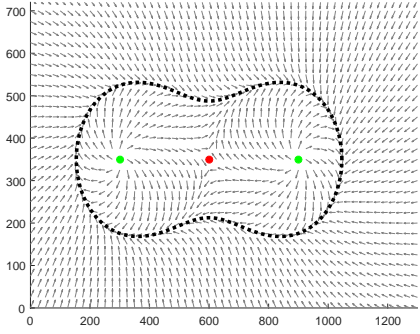}    
		\caption{The GVF for a Cassini oval.}
		\label{fig:lem-xy-field}
	\end{center}
\end{figure}

Note that the conditions of Corollary~\ref{cor:almost-all+} always hold for \emph{strictly convex} function $\vp(x,y)$ (that is, $H(x,y)>0$ at any point) such that
since the critical point (if it exists) is the unique point of \emph{global minimum}. Therefore, at this point $(x_*,y_*)$ one has $\vp(x_*,y_*)<0$ and thus $e<0$, which implies that $eH<0$.
This is exemplified by the function
\ben
\vp(x,y)=\frac{(x-x_0)^2}{a^2}+\frac{(y-y_0)^2}{b^2}-1,
\een
defining the elliptic path. Fig.~\ref{fig:ell-xy-field} demonstrates the corresponding GVF with the unique critical point at the center of ellipse. This critical point is ``repulsive'' and no trajectory of~\eqref{eq:sys-md} converges to it. Fig.~\ref{fig:lem-xy-field} illustrates the GVF for the Cassini oval, being the zero level set of the non-convex function
\ben
(x-x_0)^4+(y-y_0)^4-2a^2((x-x_0)^2-(y-y_0)^2)+a^4-b^4.
\een
The corresponding set $\mathcal C_0$ consists of two ``locus points'' $(x_0\pm a,y_0)$ and the ``center'' $(x_0,y_0)$. In
all these points $e<0$. At the locus points one has $H>0$, whereas at the center $H$ has one positive and one negative eigenvalue, so the robot potentially can be ``trapped'' at the center, but
the set of corresponding initial conditions has zero measure.

\section{The Path Following Controller Design}\label{sec:algor}

In this section, we design the controller, steering the robot to the GVF~\eqref{eq:md}. This controller
uses the GVF $\bar m_d(x,y)$ at the current robot's position and the function $\omega_d(x,y,\alpha)$, measuring the GVF's ``rotation rate'' along the robot's trajectory.

\subsection{Preliminaries}

We start with the following technical lemma.
\begin{lem}
Let $\bar m_d(t)=\bar m_d(x(t),y(t))$ stand for the GVF along a trajectory of the robot. Then its derivative $\dot{\bar m}_d(t)$ is
\be\label{eq:omega-d}
\dot m_d(t)=-\omega_d(x(t),y(t),\alpha(t))E\bar m_d(t),
\ee
where $\omega_d:\r^3\to\r$ is continuous and uniquely determined by the functions $\vp(\cdot)$, $\psi(\cdot)$ and the constant $k_n$ from~\eqref{eq:v}.
\end{lem}
\begin{IEEEproof}
Differentiating the equality $|\bar m_d(t)|^2=1$, one shows that $\dot{\bar m}_d(t)^{\top}{\bar m}_d(t)=0$, that is, $\dot{\bar m}_d(t)\perp\bar m_d(t)$.
Therefore, the vector $\dot{\bar m}_d(t)$ is proportional to the unit vector $Em_d(t)$, so that $\dot{\bar m}_d(t)=-\omega_d(t)E\bar m_d(t)$ and the scalar multiplier $\omega_d(t)$
can be found from $\omega_d(t)=-\dot{\bar m}_d(t)^{\top}E\bar m_d(t)$. It remains to prove that $\omega_d(t)$ in fact depends only on the trajectory $(x(t),y(t),\alpha(t))$, and this dependence is continuous.

Introducing the vector field~\eqref{eq:v} $\bar v(t)=\bar v(x(t),y(t))$ and the tracking error $e(t)=e(x(t),y(t))$ along the robot's trajectory, a straightforward computation shows that
\be\label{eq:mddot}
\begin{gathered}
{\dot{\bar{m}}_d}=\frac{d}{dt}\frac{\bar{v}}{\|\bar{v}\|}=\left(\frac{I_2}{\|\bar{v}\|}
-\frac{\bar{v} \bar{v}^{\top}}{\|\bar{v}\|^3}\right)\dot{\bar{v}},\\
\dot{\bar{v}}=
u_r[E-k_n e I_2] H(x,y) \bar{m}(\alpha)-k_n\dot e\,\bar{n}(x,y),\\
\dot e=u_r\psi'(\vp(x,y))\,\bar n(x,y)^{\top}\bar m(\alpha).\\
\end{gathered}
\ee
Here $I_2$ is $2\times 2$ identity matrix and $H$ is the Hessian~\eqref{eq:hessian}.
Since $\vp\in C^2$, ${\dot{\bar{m}}_d}$ and $\bar m_d$ continuously depend on the triple $(x(t),y(t),\alpha(t))$, the same holds for $\omega_d=-\dot{\bar m}_d^{\top}E\bar m_d$.
\end{IEEEproof}

To clarify the meaning of the function $\omega_d$, suppose for the moment that the robot's speed is $u_r=1$. If the robot is moving strictly along the integral curves of the GVF, then $\omega_d$ is the signed curvature of the robot's trajectory at its current position. In general, $\omega_d$ can be treated as the ``desired'' curvature of the robot's trajectory, which may differ from its real curvature.

\subsection{The algorithm of GVF steering}

As was discussed in the foregoing, the idea of the path following algorithm is to steer the robot's orientation along the guiding vector field $\bar m_d=\bar m_d(x,y)$.
We introduce the directed angle $\delta=\delta(x,y,\alpha)\in (-\pi;\pi]$ between $\bar m_d$ and $\bar m$ (see Fig.~\ref{fig:1}). The function $\delta$
is thus defined at any point $(x,y,\alpha)$ and $C^1$-smooth at the points where $\bar m(\alpha)\ne-\bar m_d(x,y)$.

The orientation vector's derivative along the trajectory is
\be\label{eq:dm}
\dot{\bar m}(t)=\frac{d}{dt}\bar m(\alpha(t))=\omega(t)\begin{bmatrix}
-\sin\alpha(t)\\
\cos\alpha(t)
\end{bmatrix}=-\omega(t)E\bar m(t).
\ee
On the other hand, $\bar m$ may be decomposed (Fig.~\ref{fig:1}) as
\begin{equation}\label{eq:decompose}
\bar m=(\cos\delta)\bar m_d-(\sin\delta)E\bar m_d=[(\cos\delta) I_2-(\sin\delta) E]\bar m_d,
\end{equation}
At any point where $\delta (t)<\pi$ and thus $\dot\delta(t)$ exists, one obtains
\[
\begin{aligned}
-\omega &E\bar m\stackrel{\eqref{eq:md}}{=}
\dot{\bar m}=-\dot\delta\underbrace{[(\sin\delta) I_2+(\cos\delta) E]\bar m_d}_{=E\bar m}+\\&+[(\cos\delta) I_2-(\sin\delta) E]\dot{\bar m}_d
\stackrel{\eqref{eq:omega-d}}{=}-\dot\delta E\bar m-\\&-\omega_d\underbrace{[(\cos\delta) I_2-(\sin\delta) E]E{\bar m}_d}_{=E\bar m}=-(\dot\delta+\omega_d)E\bar m,
\end{aligned}
\]
entailing the following principal relation between $\delta$, $\omega$ and $\omega_d$
\begin{equation}\label{eq:delta-dot}
\dot\delta=\omega-\omega_d.
\end{equation}

Furthermore, at \emph{any} time the following equality is valid
\begin{equation}\label{eq:delta-dot1}
\frac{d}{dt}\sin\delta=-\frac{d}{dt}\bar m^{\top}E\bar m_d=(\omega-\omega_d)\cos\delta.
\end{equation}
Supposing that $\delta(t_0)=\pi$ and $\omega(t)-\omega_d(t)\le 0$ for $t\approx t_0$, equation~\eqref{eq:delta-dot1} entails that $\sin\delta\ge 0$ and hence
$\delta=\frac{\pi}{2}+\arcsin(\sin\delta)$ for $t\in [t_0;t_0+\ve)$, where $\ve>0$ is sufficiently small.
In this situation, the function $\delta(x(t),y(t),\alpha(t))$ has the right
derivative\footnote{By definition, the right derivative of a function $f(t)$ at $t=t_0$ (written $f'(t_0+0)$ or $D^+f(t_0)$) is defined by
$D^+f(t_0)=\lim\limits_{t\to t_0+0}\frac{f(t)-f(t_0)}{t-t_0}$.} $\dot\delta=D^+\delta$,
satisfying~\eqref{eq:delta-dot} as $t\in [t_0;t_0+\ve)$.

We are now in a position to describe our path-following algorithm, employing the GVF ``rotation rate'' $\omega_d$ and the angle of discrepancy between the GVF and the robot's orientation $\delta$
\begin{equation}\label{eq:algorithm0}
\boxed{
\omega(t)=\omega_d(x(t),y(t),\alpha(t))-k_{\delta}\delta(x(t),y(t),\alpha(t)).}
\end{equation}
Here $k_{\delta}>0$ is a constant, determining the convergence rate.

When $\delta(x(t),y(t),\alpha(t))<\pi$, equality~\eqref{eq:delta-dot} holds and thus
\begin{equation}\label{eq:algorithm0-delta}
\boxed{
\dot\delta=\omega-\omega_d=-k_{\delta}\delta.}
\end{equation}
Furthermore, even for $\delta(x(0),y(0),\alpha(0))=\pi$ one has $\omega-\omega_d=-k_{\delta}\delta<0$ as $t\approx t_0$ and hence \eqref{eq:algorithm0-delta} retains its
validity at $t=0$, treating $\dot\delta$ as the right derivative $D^+f$.
Thus, considering $\dot\delta$ as a new control input, the algorithm~\eqref{eq:delta-dot} is equivalent to a very simple proportional
controller~\eqref{eq:algorithm0-delta}, providing, in particular, that $\delta(x(t),y(t),\alpha(t))<\pi\,\forall t>0$.

\subsection{Local existence and convergence of the solutions}\label{subsec:solutions}

In this subsection we examine the properties of the solutions of the closed-loop system~\eqref{eq:robot},~\eqref{eq:algorithm0}, rewritten as follows
\be\label{eq:closed-loop}
\begin{gathered}
\dot x(t) = u_r\cos\alpha(t),\quad \dot y(t)=u_r\sin\alpha(t),\\
\dot\alpha(t)=\omega_d(x(t),y(t),\alpha(t))-k_{\delta}\delta(x(t),y(t),\alpha(t)).
\end{gathered}
\ee

The right-hand side of \eqref{eq:closed-loop} is continuous at any point $(x_0,y_0,\alpha_0)$, where $\bar n(x_0,y_0)\ne 0$ and
$\delta_0=\delta(x_0,y_0,\alpha_0)<\pi$. However, the discontinuity at the points where $\delta_0=\pi$ makes the
usual existence theorem~\cite{Khalil} inapplicable. To avoid this problem, we consider the equivalent ``augmented'' system
\be\label{eq:closed-loop1}
\begin{gathered}
\dot x(t) = u_r\cos\alpha(t),\quad \dot y(t)=u_r\sin\alpha(t),\\
\dot\alpha(t)=\omega_d(x(t),y(t),\alpha(t))-k_{\delta}\delta_*(t),\\
\dot\delta_*(t)=-k_{\delta}\delta_*(t).
\end{gathered}
\ee
As was discussed in the previous subsection, any solution of \eqref{eq:closed-loop} satisfies~\eqref{eq:closed-loop1} with
$\delta_*(t)=\delta(x(t),y(t),\alpha(t))$, and vice versa: choosing a solution $(x(t),y(t),\alpha(t),\delta_*(t))$, where
$\delta_*(0)=\delta(x(0),y(0),\alpha(0))\in (-\pi;\pi]$, one has $\delta_*(t)=\delta(x(t),y(t),z(t))$ for any $t\ge 0$
due to~\eqref{eq:algorithm0-delta}. Unlike~\eqref{eq:closed-loop}, the right-hand side of \eqref{eq:closed-loop1}
is a $C^1$-smooth function of $(x,y,\alpha,\delta_*)$ at any point where $\bar n(x,y)\ne 0$. The
standard existence and uniqueness theorem\cite{Khalil} implies the following lemma.
\begin{lem}\label{lem:exist}
For any point $\zeta_0=(x_0,y_0,\alpha_0)$, such that $\bar n(x_0,y_0)\ne 0$, there exists the unique solution
$\zeta(t)=(x(t),y(t),\alpha(t))$ with $\zeta(0)=\zeta_0$. Extending this solution to the maximal existence interval $[0;t_*)$, one either
has $t_*=\infty$ or $t_*<\infty$ and $(x(t),y(t))\xrightarrow[t\to t_*]{} (x_*,y_*)\in\mathcal C_0$.
\end{lem}
\begin{IEEEproof}
Reducing the Cauchy problem for the closed-loop system~\eqref{eq:closed-loop} to the Cauchy problem for~\eqref{eq:closed-loop1}, one shows that the solution exists locally and is unique~\cite{Khalil}.
Let its maximally prolongable solution $(x(t),y(t),\alpha(t))$ be defined on $\Delta_*\dfb [0;t_*)$ with $t_*<\infty$. Since $|\dot{\bar r}(t)|=u_r$, the limit exists
$$
\bar r_*=\lim_{t\to t_*} \bar r(t)=\bar r(0)+\int_0^{t_*}\dot{\bar r}(t)dt.
$$
We are going to show that $\bar r_*\in\mathcal C_0$, i.e. $\bar n(\bar r_*)=0$.
Suppose, on the contrary, that $|\bar n(\bar r_*)|>0$; therefore, $|\bar v(x(t),y(t))|$ is \emph{uniformly} positive on $\Delta_*$. Using~\eqref{eq:mddot}, this implies that $\dot{\bar m}_d(t)$, and hence $\omega_d(x(t),y(t),\alpha(t))$ and $\omega(t)$ are uniformly bounded on $\Delta_*$. Thus there exists the finite limit
$$
\alpha_*=\lim_{t\to t_*} \alpha(t)=\alpha(0)+\int_0^{t_*}\omega(t)dt,
$$
enabling one to define the solution at $t=t_*$ and then to prolong it to $[t_*,t_*+\ve)$, i.e. the solution is not maximally prolonged. This contradiction proves that $\bar n(\bar r_*)=0$.
\end{IEEEproof}

Note that if $\delta(0)=0$, that is, the robot was perfectly oriented along the GVF at the starting moment,~\eqref{eq:delta-dot} implies that $\delta(t)\equiv 0$ so that the robot
follows the integral curve of the GVF. As was shown in Section~\ref{sec:field}, in this ``ideal'' situation the robot either approaches the desired path $\mathcal P$ or is driven to one of the critical points. The latter situation is practically impossible if the conditions of Corollary~\ref{cor:almost-all+} are valid since the set of integral curves, leading to the set $\mathcal C_0$, has zero measure.

One may consider~\eqref{eq:closed-loop} as a system with \emph{slow-fast} dynamics. Informally, the controller~\eqref{eq:algorithm0-delta} provides the exponential convergence of the robot to an integral curve of the GVF; after this ``fast'' transient process, the robot ``slowly'' follows this integral curve and approaches the desired trajectory, unless it is ``trapped'' in a critical point. Ignoring the ``fast dynamics'', one may suppose that the statement of Lemma~\ref{lem:converge1} remains valid for a general solution of the system~\eqref{eq:closed-loop}.
This argument, however, is not mathematically rigorous. Recalling the proofs of Lemmas~\ref{lem:converge} and \ref{lem:converge1}, one may notice that the central argument was the \emph{non-increasing} property of the Lyapunov function~\eqref{eq:V-lyap}. Although the deviation of the robot from the integral curve exponentially decreases due to~\eqref{eq:algorithm0-delta}, the non-increasing property, in general, fails; as will be shown below, even if the robot is positioned very close to the desired path, the tracking error may increase. However, this effect does not destroy the dichotomy property (any solution converges either to $\mathcal P$ or to $\mathcal C_0$) under the following assumption, which usually holds in practice, being  valid e.g. if $\vp(\cdot)$ is a polynomial.
\begin{assum}\label{ass:growth}
There exist $\theta\in (0;k_{\delta})$ and $C>0$ such that
\be\label{eq:n-bound}
|\bar n(\bar r)|\le Ce^{\theta|\bar r|}\quad\text{as $|\bar r|\to\infty$}.
\ee
\end{assum}
The latter condition can be relaxed in the case of a closed path; however, we adopt it to consider both bounded and unbounded paths in a unified way. Note that if $\vp(x,y)$ is a polynomial function, then $k_{\delta}>0$ can be arbitrarily small. 

Henceforth all Assumptions~\ref{ass:non-degen}-\ref{ass:growth} are supposed to be valid.

The following theorem, extending Lemma~\ref{lem:converge1} to the case where $\delta(0)\ne 0$, is our first main result.
\begin{thm}\label{thm:main1}
For any maximally prolonged solution $(x(t),y(t),\alpha(t))$ of~\eqref{eq:closed-loop}, defined for $t\in [0;t_*)$, one of the following statements holds:
\begin{enumerate}
\item either $t_*=\infty$ and $\dist(\bar r(t),\mathcal P)\xrightarrow[t\to\infty]{} 0$, or
\item $\dist(\bar r(t),\mathcal C_0)\to 0$ as $t\to t_*$.
\end{enumerate}
\end{thm}

In other words, for any initial condition the algorithm drives the robot to either the desired path $\mathcal P$ or $\mathcal C_0$.
\begin{IEEEproof}
In the case where $t_*=\infty$ statement~2 holds due to Lemma~\ref{lem:exist}. Suppose that $t_*=\infty$.
Differentiating the function~\eqref{eq:V-lyap} along the trajectories, it can be shown that
\be\label{eq:aux1}
\begin{split}
\dot V&=u_re\psi'(\psi^{-1}(e))\bar n^{\top}\bar m\stackrel{\eqref{eq:decompose}}{=}\\
&\stackrel{\eqref{eq:decompose}}{=}u_re\psi'(\psi^{-1}(e))\bar n^{\top}\left[\bar m_d\,\cos\delta-E\bar m_d\,\sin\delta\right]\stackrel{\eqref{eq:md}}{=}\\
&\stackrel{\eqref{eq:md}}{=}\frac{u_re\psi'(\psi^{-1}(e))}{|v|}\bar n^{\top}\left[\bar v\,\cos\delta-E\bar v\,\sin\delta\right]\stackrel{\eqref{eq:v}}{=}\\
&\stackrel{\eqref{eq:v}}{=}\frac{u_re\psi'(\psi^{-1}(e))}{\sqrt{1+k_n^2e^2}}|\bar n|(-k_ne\cos\delta+\sin\delta)=\\
&=\Phi(e)|\bar n|(-k_ne\cos\delta+\sin\delta).
\end{split}
\ee
Here $\Phi(e)$ denotes the bounded, in view of~\eqref{eq:sup}, function
\be\label{eq:PPhi}
\Phi(e)\dfb\frac{u_re\psi'(\psi^{-1}(e))}{\sqrt{1+k_n^2e^2}}.
\ee
Since $|\sin\delta|\le|\delta|$, Assumption~\ref{ass:growth} entails that
$\int_0^{\infty}|\Phi(e(t))\bar n(t)\sin\delta(t)|\,dt<\infty$. Notice now that $(-\Phi(e)e)\le 0$ and $\cos\delta(t)>0$ as $t$ becomes sufficiently large.
Thus the integral $I=\int_0^{\infty}(-e\Phi(e)|\bar n|)\cos\delta\,dt$ exists, being either finite or equal to $-\infty$. This implies, thanks to~\eqref{eq:aux1}, the existence of
$\int_0^{\infty}\dot V\,dt=\lim\limits_{t\to+\infty}V(t)-V(0)$. Since $V\ge 0$, one has $I>-\infty$ and therefore
there exists the limit $e_*=\lim_{t\to\infty}|e(x(t),y(t))|$. If $e_*=0$, then statement~1 holds due to Assumption~\ref{ass:non-degen}. Otherwise, $e\Phi(e)$ is uniformly positive and thus, recalling that $\cos\delta(t)\to 1$ as $t\to\infty$, one obtains that $\int_0^{\infty}|\bar n(x(t),y(t))|dt<\infty$, which implies statement~2 thanks to Lemma~\ref{lem:n}.
\end{IEEEproof}

\begin{cor}\label{cor:no-critical}
If $\mathcal C_0=\emptyset$, then for any initial condition the solution of~\eqref{eq:closed-loop} is infinitely prolongable and the algorithm solves the path following problem $\dist(\bar r(t),\mathcal P)\xrightarrow[t\to\infty]{} 0$.
\end{cor}

Corollary~\ref{cor:no-critical} is applicable to linear mappings $\vp(x,y)=ax+by+c$ (with $|a|+|b|\ne 0$)
and many other functions, e.g. $\vp(x,y)=y+f(x)$. These functions, however, usually correspond to \emph{unbounded}
desired curves, whereas for closed paths the GVF $\bar m_d$ is usually not globally defined.

The experiments show that under the assumptions of Corollary~\ref{cor:almost-all+} the robot always ``evades''
the finite set of critical points and converges to the desired trajectory. This looks very natural since after very fast transient dynamics the robot ``almost precisely'' follows some integral curve, which leads to $\mathcal P$ ``almost surely''. We formulate the following hypothesis.

\textbf{Hypothesis.} \emph{Under the assumptions of Corollary~\ref{cor:almost-all+}, for almost all initial conditions $(x(0),y(0),\alpha(0))$ the robot's trajectory $(x(t),y(t))$ converges to the desired path $\mathcal P$.}

Whereas the proof of this hypothesis remains a challenging problem, it is possible to guarantee the global existence of the solutions and their convergence to the desired path in some broad \emph{invariant set}, free of the critical points. The corresponding result, which does not rely on the assumptions of Corollary~\ref{cor:almost-all+},
is established in the next subsection.

\subsection{An invariant set, free of critical points}

In this subsection we give a sufficient condition, guaranteeing that a solution of~\eqref{eq:closed-loop} does not converge to $\mathcal C_0$. This criterion requires the initial condition $(x(0),y(0),\alpha(0))$ to belong to some \emph{invariant} set, free of critical points. Similar restrictions arise in
most of the path following algorithms; for example, in the projection-based algorithms the convergence can be rigorously proved only in some region of attraction where the projection to the desired curve is well defined~\cite{Matveev2013973}.

Assumptions~\ref{ass:non-degen} and~\ref{ass:positive} imply the uniform positivity of the error on
$\mathcal C_0$, that is, the following inequality holds
\be\label{eq:ec}
e_c=\inf\{|e(x,y)|:(x,y)\in\mathcal C_0\}>0.
\ee
Consider the following set
\be\label{eq:M}
\mathcal M=\left\{(x,y,\alpha):\bar n\ne 0,|\delta|<\arctan(k_ne_c),|e|<e_c\right\}.
\ee
Recall that $k_n$ is the constant parameter from~\eqref{eq:v} and $\delta=\delta(x,y,\alpha)$ is the angle between the robot's heading and the vector field direction (see Fig.~\ref{fig:1}). By definition,
$\mathcal M\cap\mathcal C_0=\emptyset$. The following lemma states that in fact $\mathcal M$ is an \emph{invariant} set, i.e. any solution starting in $\mathcal M$ remains there.
\begin{thm}\label{thm:main2}
Any solution of~\eqref{eq:closed-loop}, starting at $(x(0),y(0),\alpha(0))\in \mathcal M$, does not leave $\mathcal M$, and is infinitely prolongable satisfying the inequality
\be\label{eq:error-estim}
|e(t)|\le\max\{|e(0)|, k_n^{-1}\tan\delta(0)\}<e_c.
\ee
For such a solution, one has $\dist(\bar r(t),\mathcal P)\xrightarrow[t\to\infty]{} 0$, i.e. the algorithm~\eqref{eq:algorithm0} solves the path following problem in $\mathcal M$.
\end{thm}
\begin{IEEEproof}
Consider a solution $(x(t),y(t),\alpha(t))$, starting at $(x(0),y(0),\alpha(0))\in\mathcal M$.
Due to~\eqref{eq:algorithm0-delta}, one has $|\delta(t)|\le|\delta(0)|<\frac{\pi}{2}\,\forall t\ge 0$, and hence
$\cos\delta(t)>0$. By noticing that $\Phi(e)e\ge 0$ and thus $|\Phi(e)|=\Phi(e)\sgn e$, one has
\ben
\begin{aligned}
-\dot V&\stackrel{\eqref{eq:aux1}}{=}-u_r|\Phi(e)|\sgn e\,|\bar n|(e\cos\delta+\sin\delta)=\\
&=u_r|\Phi(e)||\bar n|(|e|+\sgn e\tan\delta)\cos\delta\ge \\
&\ge u_r|\Phi(e)|\,|\bar n|(|e|-|\tan\delta|)\cos\delta.
\end{aligned}
\een
In particular, $\dot V\le 0$ whenever $|e|\ge|\tan\delta|$.

Notice that if $|e(t)|\ge|\tan\delta(t)|$ for any $t$ (where the solution exists), then, obviously
$|e(t)|\le |e(0)|$ and thus~\eqref{eq:error-estim} holds. Suppose now that $|e(t_0)|<|\tan\delta(t_0)|$ at some $t_0\ge 0$. We are going to show that $|e(t)|\le |\tan\delta(t_0)|\le |\tan\delta(0)|$ for any $t\ge t_0$.
Indeed, had we $|e(t_1)|>|\tan\delta(t_0)|$ at some point $t_1$, there would exist $t_*<t_1$ such that $|e(t_*)|=|\tan\delta(t_0)|$ and
$|e(t)|>|\tan\delta(t_0)|$ as $t\in (t_*;t_1]$.
Using~\eqref{eq:algorithm0-delta}, it can be easily shown that $|\tan\delta(t_0)|\ge |\tan\delta(t)|$ for $t\ge t_0$,
and hence $|e(t)|>|\tan\delta(t)|$ is non-increasing when $t\in (t_*;t_1]$, which contradicts the assumption
that $|e(t_1)|>|\tan\delta(t_0)|=|e(t_*)|$. We have proved that in both cases 1) and 2) the inequality~\eqref{eq:error-estim} holds at any point where the solution exists; thus the solution stays in $\mathcal M$. By definition~\eqref{eq:ec} of $e_c$, the vector $\bar r(t)$ cannot converge to $\mathcal C_0$ in finite or infinite time, i.e. for the considered solution the statement in Theorem~\ref{thm:main1} holds.
\end{IEEEproof}

\begin{remark}\label{rem:deviation1}
The condition $(x(0),y(0),\alpha(0))\in \mathcal M$ restricts the robot to be ``properly'' headed
in the sense that
\be\label{eq:deviation}
|\delta(x(0),y(0),\alpha(0))|<\arctan(k_ne_c)<\pi/2.
\ee
Since $e_c>0$, \eqref{eq:deviation} is valid for sufficiently large $k_n$ whenever $|\delta|<\pi/2$. In other words, Theorem~\ref{thm:main2} guarantees convergence to the path
from any starting position with $|e(0)|<e_c$ and $|\delta(0)|<\pi/2$, choosing large $k_n$. Furthermore, if the desired path $\mathcal P$ is \emph{closed} and
the direction of circulation along it is unimportant, the condition $|\delta(0)|<\pi/2$ can be provided by reverting the vector field $\bar m_d\mapsto -\bar m_d$
(which corresponds to the replacement of $\vp\mapsto -\vp$ and $|\delta|\mapsto \pi-|\delta|$) unless $\delta(0)=\pm\pi/2$ (in practice it is never possible).
\end{remark}
\begin{remark}\label{rem:deviation}
Even if~\eqref{eq:deviation} is violated at the starting time $t=0$, it obviously holds when $t>t_0=k_{\delta}^{-1}[\ln|\delta(0)|-\ln\arctan(k_ne_c)]$
thanks to~\eqref{eq:algorithm0-delta}. Thus, if one is able to prove that the solution is prolongable up to $t_0$ and $|e(x(t_0),y(t_0))|<e_c$, Theorem~\ref{thm:main2}
provides the convergence of the robot's position to the desired path.
\end{remark}

Remark~\ref{rem:deviation} suggests the way to relax the restriction on the initial robot's orientation~\eqref{eq:deviation}. Since the robot moves at the constant speed  $u_r>0$, it covers the distance $u_rt_0$ until its orientation satisfies~\eqref{eq:deviation}. The path following is guaranteed if
$u_rt_0$ is less than the \emph{viability distance} of the initial position, i.e. the distance from it to the set where $|e|\ge e_c$.
\begin{defn}
Given a point $\bar r_0=(x_0,y_0)$ with $|e(\bar r_0)|<e_c$, the number $d_0=\inf\{\dist(\bar r_0,\bar r):e(\bar r)\ge e_c\}>0$ is said to be its~\emph{viability distance}.
\end{defn}

Theorem~\ref{thm:main2} and Remark~\ref{rem:deviation} yield in the following.
\begin{cor}\label{cor:viability}
Let the initial position of the robot $\bar r(0)=(x(0),y(0))$ with $e(\bar r(0))<e_c$ have the viability distance $d_0>0$. If this viability distance satisfies the condition
\be\label{eq:viability-1}
d_0>\frac{u_r}{k_{\delta}}\ln\frac{|\delta(x(0),y(0),\alpha(0))|}{\arctan(k_ne_c)},
\ee
then the solution of~\eqref{eq:closed-loop} gets into the set $\mathcal M$ in finite time, and is prolongable up to $\infty$ satisfying $\dist(\bar r(t),\mathcal P)\xrightarrow[t\to\infty]{} 0$.
\end{cor}

Note that~\eqref{eq:viability-1} holds for any $\alpha(0)$ under the assumption
\be\label{eq:viability-2}
d_0>\frac{u_r}{k_{\delta}}\ln\frac{\pi}{\arctan(k_ne_c)}.
\ee
The condition~\eqref{eq:viability-2} gives an estimate for the region, starting in which the robot necessarily converges to the desired path $\mathcal P$.
Taking the original orientation relative to the field into account, this estimate can be tightened by using~\eqref{eq:viability-1}.
Typically $d_0$ is \emph{uniformly} positive in the vicinity of $\mathcal P$ (this holds, for instance, when $\mathcal P$ is a closed curve). Thus~\eqref{eq:viability-2} guarantees the algorithm
to converge in the vicinity of $\mathcal P$, choosing $k_{\delta}/u_r$ sufficiently large. As was mentioned, practical experiments with natural trajectories, satisfying the assumptions of Corollary~\ref{cor:almost-all+}, show that the robot is always attracted to the desired path, although the mathematical proof of this remains a non-trivial task.
\begin{remark} Although explicit calculation of the viability distance is complicated, conditions~\eqref{eq:viability-1} and \eqref{eq:viability-2} in fact require only its \emph{lower} bound.
Such a bound can be explicitly obtained, for instance, if the error is Lipschitz
\be\label{eq:e-nabla}
|\nabla e(\bar r)|=\psi'(\vp(\bar r))|\bar n(\bar r)|\le c=const\quad\forall x,y.
\ee
If the condition~\eqref{eq:e-nabla} holds, then $|e(\bar r)-e(\bar r_0)|\le c|\bar r-\bar r_0|$ and hence the viability distance of $\bar r_0$ is estimated as follows
$$
d_0\ge (e_c-e(x_0,y_0))/c.
$$
\end{remark}

Condition~\eqref{eq:e-nabla} can often be provided under an appropriate choice of $\psi(\cdot)$. For instance, when $\vp(x,y)=(x-x_0)^2+(y-y_0)^2-R^2$ (circular path), one can choose
$\psi(s)=\arctan s$ so that $\psi'(s)=1/(1+s^2)$. More generally, if $|\vp(\bar r)|\ge C_1|\bar r|^{\beta}$ and $|\nabla\vp(r)|\le C_2|\bar r|^{\gamma}$ as $|\bar r|\to\infty$, then
$\nabla e$ is globally bounded; we choose $\psi(s)=\arctan (s^p)$, where $p>0$ is sufficiently large so that $p-1+\gamma\le 2\beta p$.

\section{Experimental validation}\label{sec:exper}

In this section the experiments with real robots are reported. In one of these experiments, the desired path is an ellipse, and the other experiments deals with the more sophisticated Cassini oval (see Section~\ref{sec:field}).

\begin{figure}
	\begin{center}
		\includegraphics[width=0.75\columnwidth]{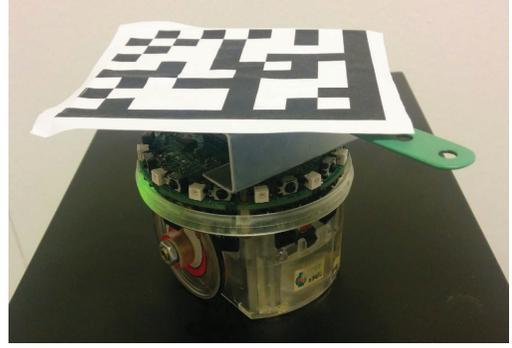}    
		\caption{The E-puck robot with marker on top, used in experiments}
		\label{fig:puck}
	\end{center}
\end{figure}

We test the results using the E-puck mobile robotic platform \cite{ePuck}. The experimental setup consists of the differential wheeled E-puck robot,  a server computer,
an overhead camera and a communication module. The robot is identified by a data matrix marker on its top (see Fig.~\ref{fig:puck}).
The position of the central point and the orientation of the marker are recognized by a vision  algorithm running at a control computer connected to the overhead camera.
The available workspace is a planar area of 2.6x2 meters covered by the image of 1280x720 Px (henceforth the acronym Px is used for pixels). For convenience we operate with pixels as coordinates. The PC runs real-time calculation of the control actions based on the pose information of the robot and computes the control inputs for the
robot. The results of computation as the desired angular and linear velocities of the robot are translated into the commands for its left and right wheels.
The commands from the control computer to the robot are sent via Bluetooth at the fixed frequency of 20 Hz.

We verify the guidance algorithm for two closed trajectories, passed by the robot in the \emph{clockwise} direction.
In both cases the forward velocity was set to $u_r = 50$ Px/s, and we set the parameters of the algorithm $k_n = 3$ and $k_{\delta} = 2$. We choose $\psi(s)=s$, so that $e(x,y)=\vp(x,y)$. Below we describe the trajectories and show the numerical data from the experiments.

\subsection{Circulation along the ellipse}

For our first experiment, we choose the elliptic trajectory, defined by the function
\begin{equation}\label{comp:ellipse}
\varphi(x, y) = k_s \left( \frac{(x - x_0 )^2}{p^2} + \frac{(y - y_0 )^2}{q^2} - R^2  \right).
\end{equation}
Obviously, the level curves of $\vp$ are ellipses. The path $\mathcal P$ corresponds to the ellipse with semiaxes $pR$ and $qR$, centered at the point $(x_0,y_0)$.

\begin{figure}[h]
	\begin{center}
		\includegraphics[width=0.8\columnwidth]{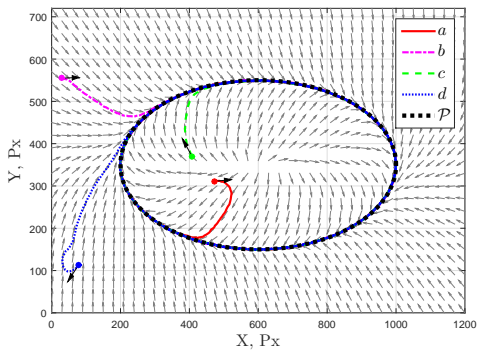}    
		\caption{Elliptical path: the vector field and robot's trajectories.}
		\label{fig:ell-xy}
	\end{center}
\end{figure}

\begin{figure}[h]
	\begin{center}
		\includegraphics[width=0.8\columnwidth]{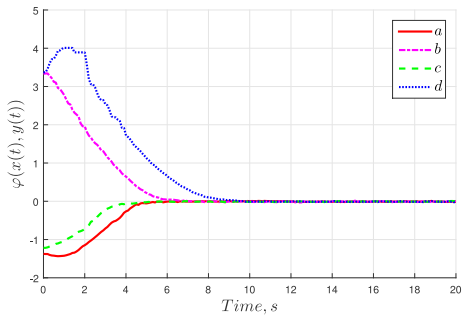}    
		\caption{Elliptical path: the dynamics of the tracking error $e=\vp(x,y)$.}
		\label{fig:ell-e}
	\end{center}
\end{figure}

For the experiment we choose $x_0=600$Px, $y_0=350$Px, $R=400$Px, and the semiaxes scale factors $p = 1$, $q = 0.5$ (see Fig.~\ref{fig:ell-xy}). To provide $\vp(x,y)\in [-5;5]$ in the working area, we choose the scaling factor $k_s = 10^{-5}$. The robots's trajectories (labeled $a$, $b$, $c$ and $d$) under four different initial conditions are shown in Fig.~\ref{fig:ell-xy}. The respective starting conditions are
\be\label{eq:initial1}
\begin{split}
a&: (x = 472, y = 311 , \alpha=0.0768 ),\\
b&: (x = \ 30 , y = 555, \alpha=0.0278), \\
c&: (x = 408, y = 369, \alpha=2.1515),\\
d&: (x = \ 78, y = 133, \alpha=4.0419)
\end{split}
\ee
(the coordinates $x,y$ are in pixels, and $\alpha$ is in radians). Fig.~\ref{fig:ell-e} illustrates the dynamics of the tracking error $\vp(x(t),y(t))$.

In this example the set of critical points is $\mathcal C_0=\{(x_0,y_0)\}$, which corresponds to $e(x_0,y_0)=-k_sR^2$ and hence $e_c=R^2$. The set $\{(x,y):|e(x,y)|<e_c\}$ consists of all points, where
\be\label{eq:exp1-domain}
0<\frac{(x-x_0)^2}{p^2}+\frac{(y-y_0)^2}{q^2}<2R^2.
\ee
Theorem~\ref{thm:main2} guarantees that starting at any of these points and~\eqref{eq:deviation} holds, the robot converges to the desired path. Using the geometry of the specific set~\eqref{eq:exp1-domain}, Remark~\ref{rem:deviation} allows to prove convergence for many other initial conditions: it is clear, for instance,
that if the robot starts from some point, lying outside the ellipse and sufficiently far from it, then it enters into the domain~\eqref{eq:exp1-domain} with
a ``proper'' orientation, satisfying~\eqref{eq:deviation}, and thus the problem of path following is solved.
The practical experiments, however, show that the robot converges to the path from \emph{any} point, except for the ellipse' center, independent of the initial robot's
orientation, whereas the mathematically rigorous proof remains an open problem.

\subsection{Circulation along the Cassini oval}

\begin{figure}[h]
	\begin{center}
		\includegraphics[width=0.8\columnwidth]{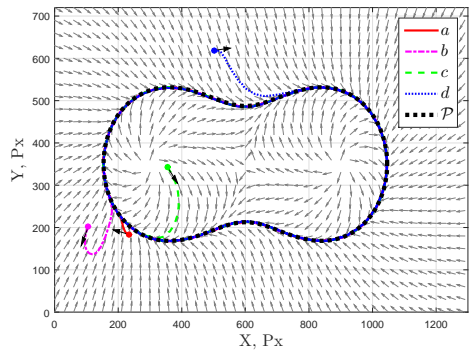}    
		\caption{Cassini oval: the vector field and robot's trajectories.}
		\label{fig:lem-xy}
	\end{center}
\end{figure}

For our second experiment the path chosen is referred to as the \emph{Cassini oval} (Fig.~\ref{fig:lem-xy}) and determined by the function
\ben
\begin{split}
\varphi(x,y) &= k_s \left[ \left( \Delta x^2 + \Delta y^2 \right)^2 - 2 q^2 \left( \Delta x^2 - \Delta y^2 \right) - p^4 + q^4 \right],\\
\Delta x &= (x - x_0 ), \ \Delta y = (y - y_0).
\end{split}
\een
We choose here $k_s = 10^{-10}$, $p = 330$Px and $q = 300$Px. The oval is centered at $(x_0,y_0)=(600,350)$Px.

\begin{figure}[h]
	\begin{center}
		\includegraphics[width=0.8\columnwidth]{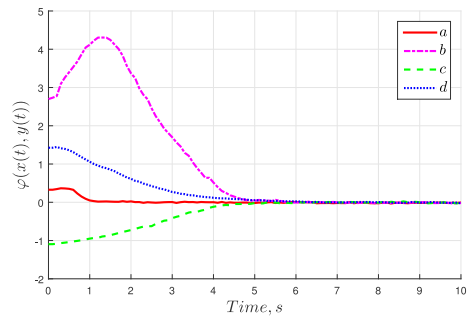}    
		\caption{Cassini oval: the dynamics of the tracking error $e=\vp(x,y)$.}
		\label{fig:lem-e}
	\end{center}
\end{figure}

Fig.~\ref{fig:lem-xy} illustrates four trajectories (labeled $a$, $b$, $c$ and $d$), corresponding to the initial conditions
\be\label{eq:initial2}
\begin{split}
a&: (x = 233, y = 184, \alpha=2.9287 ),\\
b&: (x = 106, y = 202, \alpha=4.2487), \\
c&: (x = 355, y = 343, \alpha=5.4071),\\
d&: (x = 503, y = 619, \alpha=0.1022).
\end{split}
\ee
(the coordinates $x,y$ are in pixels, and $\alpha$ is in radians). The corresponding tracking errors are displayed in Fig.~\ref{fig:lem-e}.

\begin{figure}[h]
	\begin{center}
		\includegraphics[width=0.8\columnwidth]{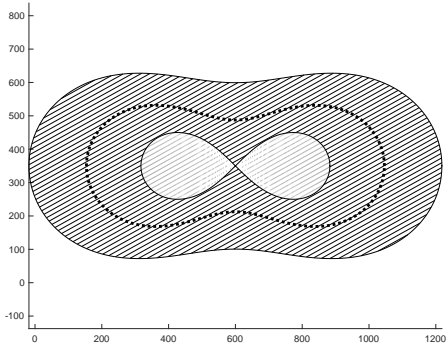}    
		\caption{Cassini oval: the set $\{(x,y):|e(x,y)|<e_c\}$.}
		\label{fig:between-lem}
	\end{center}
\end{figure}

As discussed in Section~\ref{sec:field}, the set of critical points is $\mathcal C_0=\{(x_0\pm q,y_0),(x_0,y_0)\}$. If $q^2<p^2<2q^2$, such inequalities being valid in our example, then
it can be checked that $e_c=|e(x_0,y_0)|=p^4-q^4$, and the set $\{(x,y):|e(x,y)|<e_c\}$ is a domain, shown in Fig.~\ref{fig:between-lem}.
Theorem~\ref{thm:main2} guarantees that starting at any of these points with the orientation satisfying~\eqref{eq:deviation}, the robot converges to the desired path.
Similar to the case of the elliptic path, Remark~\ref{rem:deviation} allows to prove convergence for many other initial conditions; experiments show that in fact the robot converges
to the desired path from any non-critical point.

\section{Discussion}\label{sec:discuss}

In this section, we compare the proposed method with the existing path following control algorithms. Potential extensions of the proposed method are also discussed.

\subsection{The GVF method vs. other path following algorithms}

It should be noticed that many approaches to path following, discussed in Introduction, are not applicable to general smooth paths of non-constant curvature and nonholonomic robots; for example, various virtual target approaches~\cite{1272868} typically require to control the robot's velocity. We compare the GVF method with two commonly used algorithms~\cite{6712082}: the projection-based line-of-sight (LOS) method~\cite{Fossen20142912,1582226} and the ``nonlinear guidance law'' (NGL)~\cite{6712082,NLGL}.

\begin{figure}[h]
	\begin{center}
		\includegraphics[width=0.75\columnwidth]{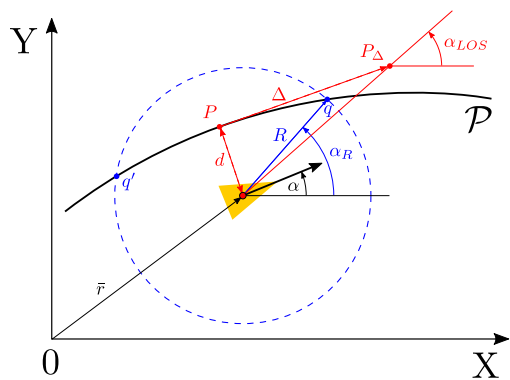}    
		\caption{Geometric objects, employed by the LOS (red) and the NGL (blue) path following algorithms}
		\label{fig:comparison_1}
	\end{center}
\end{figure}

The LOS approach assumes the existence of the unique \emph{projection} of the robot location $\bar{r}$ onto the path $\mathcal{P}$, that is, the point $P$ closest to $\bar r$ in the Euclidean metric. The distance to the path $d=d(\bar r,P)$ serves as the tracking (or ``cross-track'') error. The desired robot's orientation $\alpha_{LOS}$ is the absolute bearing to the unique point $P_{\Delta}$ (Fig.~\ref{fig:comparison_1}), such that the vector $\overrightarrow{PP_{\Delta}}$ has length $\Delta>0$
and is tangent to $\mathcal P$ at the projection point. The direction of this vector is the desired direction of path following (in our example, the robot circulates along the path clockwise).
The controller~\eqref{eq:algorithm0} is replaced by
\begin{equation*}\label{comp:los}
\omega_{LOS} = c({P}) u_r - k_{LOS}\left(\alpha - \alpha_{LOS}\right),
\end{equation*}
where $k_{LOS}>0$ is a constant gain and $c({P})$ is the curvature of the curve $\mathcal{P}$ at the projection point ${P}$. The parameters of the algorithm are the \emph{lookahead distance} $\Delta$ and the gain $k_{LOS}$.

The second path following algorithm to be compared with the GVF method is the ``nonlinear guidance law'' (NGL)~\cite{6712082,NLGL}. At the current robot position $\bar{r}$, draw a circle of radius $R$. The circle intersects the path $\mathcal{P}$ at two points $q$, $q'$. It is supposed that the robot moves sufficiently close
to the desired path and the direction of the path following is fixed, so it is always possible to choose the intersection point, lying \emph{ahead} of the robot on the path $\mathcal P$ (in Fig.~\ref{fig:comparison_1}, this is
point $q$). The absolute bearing $\alpha_R$ to this point is the desired direction of motion.
The angular velocity controller is designed as
\begin{equation*}
\omega_{R} = -k_{R}\left(\alpha - \alpha_{R}\right),
\end{equation*}
The constants $R>0, k_R>0$ are the algorithm's parameters.

\begin{figure}[h]
	\begin{center}
		\includegraphics[width=\columnwidth]{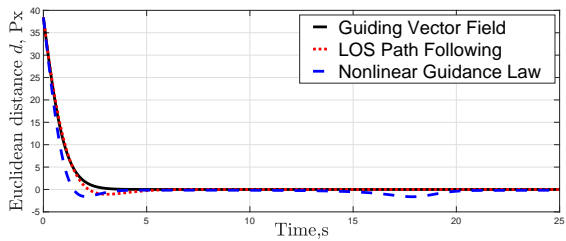}    
		\caption{The dynamic of the euclidean distance $d$}
		\label{fig:comparison_2}
	\end{center}
\end{figure}

For the sake of comparison, we will use the same elliptic path, determined by the function~\eqref{comp:ellipse} and the same parameters of the GVF algorithm as in Subsect.~\ref{sec:exper}-A.
The initial position and heading of the robot are
\begin{equation*}
x = 200; y = 450; \alpha = 0.0278.
\end{equation*}
It is clear that any path following controller is very sensitive to the parameter choice, and varying the parameters one can always make the convergence rate faster or slower. To compare the behavior of different algorithms, one thus needs to choose parameters, providing approximately the same convergence rate; in our situation, the dynamics of the Euclidean distance $d(t)=\dist(\bar r(t),\mathcal P)$ is similar to the GVF method, choosing $k_{LOS} = 2$ and $\Delta = 70$ in the LOS method and $k_{R} = 2$ and $R = 40$ in the NGL algorithm (Fig.~\ref{fig:comparison_2}).
As one can see, the GVF method gives much better transient behavior than the other algorithms, which have visible overshoots. To eliminate the overshoot of the LOS method, one has to increase the lookahead distance, but in this case the convergence to the desired path becomes slower. The NGL method has the largest overshoot; using this method,
it is also impossible to eliminate completely the tracking error since the controller has no information about the changing curvature of the path. To decrease the overshoot, one has to increase the radius $R$; this however amplifies the oscillations in the tracking error $d$. Thus the GVF method demonstrates better performance than the considered alternative methods.

\subsection{The previous works on GVF algorithms}

The vector-field path following algorithms have been mainly considered for straight lines and circular paths~\cite{4252175,6712082}.
The GVF designed for these paths is the same as constructed in Section~III; the only difference is the path following controller:
unlike~\cite{4252175}, in this paper we do not consider sliding-mode algorithms and confine ourselves to simple linear controllers.
However, a general approach to the Lyapunov-based GVF design has been suggested in~\cite{lawrence2008lyapunov}. The desired path is supposed to be the set of global minima
of some predefined Lyapunov function; e.g. for a path~\eqref{eq:phi}, our usual Lyapunov function~\eqref{eq:V-lyap} can be chosen. Departing from this Lyapunov function, a broad class of GVF has been suggested, including~\eqref{eq:md} as a special case. The paper~\cite{lawrence2008lyapunov} also offers a methodology of proving the convergence of integral curves to the desired path (considered in our Section~\ref{sec:field}) and design of path following controllers.
For the case of unicycle-like robot the control algorithm, suggested in~\cite[Eq.~(35)]{lawrence2008lyapunov} is equivalent to our controller~\eqref{eq:algorithm0-delta}. Describing a general framework, the paper~\cite{lawrence2008lyapunov} focuses however on some specific problems of flight control, namely, following a \emph{loiter circle} (planar circle at the fixed altitude) or some other planar path, which can be obtained from a loiter circle by a smooth ``warping'' transformation, preserving the vector field's properties.
Many assumptions adopted in~\cite{lawrence2008lyapunov} are restrictive, e.g. the Lyapunov function should be radially unbounded, excluding straight lines and other unbounded paths from the consideration.
The analysis of the closed-loop system in~\cite{lawrence2008lyapunov} is not fully rigorous as it completely
ignores the problem of the solution existence up to $\infty$, which becomes non-trivial when $\mathcal C_0\ne\emptyset$.

Unlike the previous papers~\cite{4252175,6712082,lawrence2008lyapunov}, we consider a general smooth path~\eqref{eq:phi}, which can be
closed or unbounded; for closed trajectories, we do not restrict the Lyapunov function to be radially unbounded. We provide a mathematically rigorous analysis of our algorithm and specify an explicit set of initial conditions, starting at which the robot evades the critical points and converges to the desired trajectory.

\subsection{Further extensions of our GVF method}

The results presented in the paper can be extended in several directions. First, the idea of GVF steering can be applied to more general models than the simplest unicycle robot, considered in this paper.

\begin{figure}
	\begin{center}
		\includegraphics[width=0.8\columnwidth]{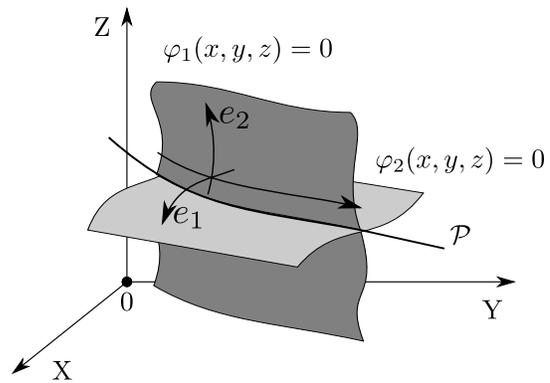}    
		\caption{The desired path as the intersection of two surfaces}
		\label{fig:3D}
	\end{center}
\end{figure}

Second, the algorithm can be extended to the three-dimensional case. Some special results in this direction
have been reported in~\cite{lawrence2008lyapunov,micnon}.
The desired curvilinear path $\mathcal{P}$ in the three dimensional space can be describing as an intersection of two surfaces (Fig.~\ref{fig:3D})
\begin{equation*}\label{3d:path}
\mathcal{P} \dfb \left\{ \left(x,y,z\right):\varphi_1(x,y,z)=0 \wedge \varphi_2(x,y,z)=0 \right\} \subset \mathbb{R}^3,
\end{equation*}
where $\varphi_i \in C^2(\mathbb{R}^3 \rightarrow \mathbb{R}),i = 1,2$.
The main design idea is that if the robot simultaneously approaches both surfaces, eventually it will be brought to the prescribed path $\mathcal{P}$. Thus, we may define the guiding vector field by 	
\begin{equation*}\label{3d:v}
\bar{v} = \bar{\tau}- k_{n1} e_1 \bar{n}_1- k_{n2} e_2 \bar{n}_2, \quad \bar{\tau} = \bar{n}_1 \times \bar{n}_2
\end{equation*}
where $\bar{n}_i=\nabla\varphi_i$ and $e_i=\vp_i(x,y,z)$ are the ``tracking errors'' ($i=1,2$). Subsequent design of the controller, steering the robot to the GVF, is
similar to the planar case (two angles are controlled instead of a single angle).
The technical analysis of this algorithm requires however  some extra assumptions of non-degeneracy and is beyond the scope of this paper; the convergence rate of the path following algorithm appears to be very sensitive to the choice of $\varphi_1$ and $\varphi_2$.

Third, the results can be extended to time-varying vector fields~\cite{lawrence2008lyapunov}, allowing to work
with some types of transformations (such as translation, rotation and scaling) of the predefined trajectory
and enabling to use the GVF approach  for stand-off tracking of slowly moving targets~\cite{7126199}.

Fourth, it is important to consider external disturbances, unavoidable in practice and leading, in particular, to the lateral drift of the robot. Some preliminary results on using draft compensators in parallel with the path following
controller are reported in the conference papers~\cite{ICRA_VF,6740857}.

\section{Conclusions}

In this paper, we offer a new algorithm for path following control of nonholonomic robots exploiting the idea of a guiding vector field. Unlike the existing results,
the desired path can be an arbitrary smooth curve in its implicit form, i.e. the zero set of a smooth function. We examine mathematical properties of the algorithm and give global conditions
for following asymptotically the desired path. The results have been experimentally validated, using the E-puck wheeled robots.

\bibliographystyle{IEEETran}
\bibliography{lib}

\end{document}